\documentclass[journal]{IEEEtran}
\usepackage{hyperref} 

\usepackage{cite}
\usepackage{amsmath}
\usepackage{graphicx}
\usepackage{url}
\usepackage{subfigure}
\usepackage{color}

\newcommand{\ud}{\mathrm{d}}
\newcommand{\eqn}[1]{(\ref{#1})}

\newcommand{\Fig}[1]{Figure~\ref{#1}}

\newcommand{\fig}[1]{Fig.~\ref{#1}}
\newcommand{\figs}[1]{Figs.~\ref{#1}}

\newcommand{\revise}[2]{#2}
\newcommand{\revisenew}[1]{#1}
\newcommand{\remark}[1]{}

\begin{document}

\title{Statistics of Group Delays in \\ Multimode Fiber with Strong Mode Coupling}

\author{Keang-Po Ho,~\IEEEmembership{Senior Member,~IEEE}, Joseph M. Kahn,~\IEEEmembership{Fellow,~IEEE}%
\thanks{Manuscript received 2011, revised 2011.}%
\thanks{The research of JMK was supported in part by National Science Foundation Grant Number ECCS-1101905 and Corning, Inc.}
\thanks{K.-P. Ho is with Silicon Image, Sunnyvale, CA 94085. (E-mail: kpho@ieee.org)}
\thanks{J. M. Kahn is with the Edward L. Ginzton Laboratory, Department of Electrical Engineering, Stanford University, Stanford, CA 94305. (E-mail: jmk@ee.stanford.edu)}
}

\maketitle

\begin{abstract}
\revise{2.1}{The modal group delays (GDs) are a key property governing the dispersion of signals propagating in a multimode fiber (MMF). 
A MMF is in the strong-coupling regime when the total length of the MMF is much greater than the correlation length over which local principal modes can be considered constant.}
In this regime, the GDs can be described as the eigenvalues of zero-trace Gaussian unitary ensemble, and the probability density function (p.d.f.) of the GDs is the eigenvalue distribution of the ensemble. 
For fibers with two to seven modes, the marginal p.d.f.~of the GDs is derived analytically. 
For fibers with a large number of modes, this p.d.f.~is shown to approach a semicircle distribution. 
In the strong-coupling regime, the delay spread is proportional to the square root of the number of independent sections, or the square root of the overall fiber length.  
\end{abstract}



\section{Introduction}

\PARstart{M}ULTIMODE fiber (MMF) is widely used in short-range optical links \cite{benner05, 802.3, koike08}, where it is often favored over single-mode fiber (SMF) because of relaxed connector alignment tolerances and reduced transceiver component costs.
\revisenew{MMF supports propagation of multiple spatial modes having
different group velocities, and thus different group delays (GDs),} an effect called modal dispersion \cite{gloge72, ghatak80}.
Even if a signal is launched into one spatial mode, bends, index imperfections and other perturbations cause the signal to couple into multiple modes \cite{kitayama80, gloge72, olshansky75, raddatz98}, making the signal subject to modal dispersion.
Modal dispersion limits current commercial MMF links to 10 Gb/s per fiber up to about 300~m long \cite{802.3, 802.3aq}, and next-generation 100 Gb/s Ethernet MMF systems use ten fibers per link \cite{802.3ba}. 
Techniques to increase the bit rate per fiber are desired.

SMF, which is free from modal dispersion, is the dominant medium for longer transmission distances. 
Emerging long-haul systems use dual-polarization quaternary phase-shift keying and coherent detection to achieve a spectral efficiency of 2 bits/s/Hz \cite{robert10, yu10}. 
Usage of higher-order modulation formats \cite{kahn04} can at least double the spectral efficiency, but further increases are expected to become increasingly difficult \cite{winzer10}, because of  limits posed by optical amplifier noise and fiber nonlinearity \cite{mitra01, essiambre10}. 
Techniques to further increase spectral efficiency are desired.

Mode-division multiplexing (MDM) in MMF \cite{berdague82, stuart00}, a form of multi-input, multi-output transmission, is a potential means to increase transmission capacity in both short- and long-distance optical networks. 
Like multipath propagation in wireless systems, the plurality of modes in MMF was long viewed as a strictly negative, bandwidth-limiting effect requiring mitigation, but is now seen as creating additional degrees of freedom in which to transmit information \cite{shah05, gasulla08, li11, ryf11}.
\revise{1.a}{Modal dispersion in MMF typically leads to a larger GD spread than that caused by chromatic dispersion. 
This GD spread determines the required cyclic prefix length in MDM systems using orthogonal frequency-division multiplexing\cite{li11} or the required number of equalizer taps in MDM systems using single-carrier modulation\cite{ryf11}. In other words, receiver complexity increases in proportion to the GD spread caused by modal dispersion.
}

Effective mitigation of modal dispersion or optimal use of MDM requires a detailed understanding of modal dispersion, especially the effect of mode coupling on the modal GDs. 
Models for mode coupling were developed more than 30 years ago \cite{gloge72, olshansky75, ghatak80}, when MMF links used spatially and temporally incoherent light-emitting diodes. 
Virtually all the models ignore phase effects, and consider only power coupling between modes. 
Power coupling models are able to qualitatively explain some observations, such as the scaling of delay spread with fiber length. 
Delay spread scales linearly with fiber length in the weak-coupling regime (e.g., short glass MMF), and with the square root of fiber length in the strong-coupling regime (e.g., plastic MMF) \cite{garito98}. 
However, most modern MMF systems use spatially and temporally coherent laser sources, and power coupling models are not able to explain certain observations, such as a sensitivity of the impulse response to launched polarization \cite{yam04}. 

SMF supports propagation in two polarizations, and polarization-mode dispersion (PMD) has long been modeled using electric field coupling models \cite{poole86, li98}. 
Field coupling models have been used to demonstrate the existence of principal states of polarization (PSPs), which have well-defined group delays to first order in frequency \cite{kogelnik02, poole86, poole97}. 
In long SMFs, polarization modes are strongly coupled. 
In this regime, the differential group delay (DGD) between the PSPs scales with the square root of fiber length  and follows a Maxwellian distribution \cite{foschini91, kogelnik02, poole97, karlsson01}. 
PSPs form the basis of techniques for optical PMD compensation in direct detection systems.

Recently, field coupling models have been extended to MMF \cite{fan05, shemirani09}.
These models explain the polarization sensitivity of mode coupling and demonstrate the existence of principal modes (PMs).
The PMs have well-defined GDs to first order in frequency, and form the basis for optical techniques to compensate modal dispersion \cite{shen05}.  
The GD differences between PMs scale linearly with fiber length in the weak-coupling regime, and with the square root of fiber length in the strong-coupling regime \cite{shemirani09}. 
To date, however, the statistical properties of the GDs, which are of particular interest in the strong-coupling regime, have not been studied.

Here, the statistics of the GDs are derived analytically for  MMF in the strong-coupling regime, considering a number of modes ranging from two to infinity\footnote{Throughout this paper, ``modes'' include both polarization and spatial degrees of freedom. 
For example, the two-mode case can describe the two polarization modes in SMF.}. 
In this regime, regardless of the number of modes or the group delays in the absence of coupling, the GDs scale with the square root of fiber length or the square root of the number of independent fiber sections, similar to PMD in SMF in the strong-coupling regime \cite{kogelnik02}.
End-to-end modal dispersion effects are described, at each single frequency, by a random complex Gaussian Hermitian matrix or Gaussian unitary ensemble \cite{mehta}.  
From such a model, the joint \revise{2.2}{probability density function} (p.d.f.)~of GDs can be derived analytically. 
Here, closed-form expressions for the GD distributions are derived for small number of modes. 
For a large number of modes, the GD distribution asymptotically approaches a semicircle distribution with a radius or upper limit equal to twice its standard deviation. 

The remaining parts of this papers are organized as follows.
Sec.~\ref{sec:rmm} describes the random matrix model for MMF propagation.
Sec.~\ref{sec:fmf} provides closed-form analyses of the GD distribution in fibers with two to seven modes.
Sec.~\ref{sec:infmode} presents asymptotic expressions for the GD distribution in the limit of a large number of  modes. 
Secs.~\ref{sec:dis} and \ref{sec:con} are discussion and conclusions, respectively.

\section{Random Matrix Model for Multimode Fibers}
\label{sec:rmm}

The propagation characteristics of a MMF, in particular, the local PMs and their GDs, can be considered invariant over a  certain correlation length.
Because bends, mechanical stresses and manufacturing tolerances induce mode coupling, the local PMs in sections separated by distance longer than the correlation length can be considered independent of each other. 
Throughout this paper, we consider the regime of strong mode coupling, where the total length of the MMF far exceeds the correlation length. 
The theory presented here is valid regardless of the actual correlation length or the modal GD profile within the correlation length.

A MMF may be divided into $K$ sections, with propagation in each section modeled as a random matrix. 
The length of each section should be at least slightly longer than the correlation length, so that the local PMs in the different sections can be considered independent. 
Although the approach used here is applicable even if each section has different properties, for convenience, we assume that all sections are statistically equivalent.

In a MMF with $D$ modes, modal propagation in the $k$th section can be modeled as a $D \times D$ matrix  $\mathcal{M}^{(k)}(\omega)$ as a function of frequency $\omega$.
Here, we are only interested in the statistical properties of the modal GD, so for simplicity, we ignore any mode-dependent gain or loss. 
With strong mode coupling, the $k$th section may be represented by the product of three $D \times D$ matrices
\begin{equation}
\mathcal{M}^{(k)}(\omega) = \mathcal{V}^{(k)} \Lambda^{(k)}(\omega) \mathcal{U}^{(k)\dagger}, k = 1, \dots, K,
\label{mmat}
\end{equation}
where $^\dagger$ denotes Hermitian transpose, $\mathcal{U}^{(k)}$ and $\mathcal{V}^{(k)}$ are random unitary matrices representing the mode coupling at the input and output, respectively, and $\Lambda^{(k)}(\omega)$ is a diagonal matrix describing the uncoupled modal GDs, i.e.,
\begin{equation}
\Lambda^{(k)}(\omega) = \mathrm{diag}\left[e^{-j \omega \tau^{(k)}_{1}}, e^{-j \omega \tau^{(k)}_{2}}, \cdots, e^{-j \omega \tau^{(k)}_{D}}\right],
\end{equation}
where $\tau^{(k)}_i$, $k = 1, \dots, K$, $i = 1, \dots, D$, are the uncoupled GDs in the sections.

\remark{The same as before, just move this paragraph earlier.}

In the absence of mode-dependent gain or loss, $\mathcal{M}^{(k)}$, $\mathcal{U}^{(k)}$ and $\mathcal{V}^{(k)}$ are all unitary matrices, such that $\mathcal{M}^{(k)}(\omega) \mathcal{M}^{(k)\dagger}(\omega) = I$, where $I$ is the identity matrix.
With strong random coupling, both $\mathcal{U}^{(k)}$ and $\mathcal{V}^{(k)}$ can be assumed to be independent random unitary matrices, such that both input and output are randomly oriented.
The model \eqn{mmat} is similar to the matrix model of PMD described in \cite{gordon00, karlsson01}. 

The model here is valid regardless of whether or not the vectors $\vec{\tau}^{(k)} = \left(\tau^{(k)}_{1}, \tau^{(k)}_{2}, \dots, \tau^{(k)}_{D}\right)$, $k = 1, \dots, K$, have the same statistical properties.
The vector $\vec{\tau}^{(k)} $ may even be a deterministic vector, identical for each section.
For convenience and without loss of generality, we assume that $\sum_i \tau^{(k)}_i = 0$, i.e., we ignore the mode-averaged delay of each section, as it does not lead to modal dispersion.

Using $|s_k \rangle$ and $|t_k \rangle$ to denote the input and output modes, respectively, we have 
\begin{equation}
|t_k \rangle = \mathcal{M}^{(k)}(\omega) |s_k \rangle
\label{inout}
\end{equation}
and $| s_k \rangle = \mathcal{M}^{(k)\dagger}(\omega) | t_k \rangle$.
Similar to the analysis of PMD \cite{poole86, poole97, gordon00}, the GDs correspond to the eigenvalues of $j \mathcal{M}^{(k)}_\omega \mathcal{M}^{(k)\dagger} $ where $ \mathcal{M}^{(k)}_\omega = \partial  \mathcal{M}^{(k)}(\omega)/\partial \omega$.
With only a single section, we may verify that
\begin{equation}
j \mathcal{M}^{(k)}_\omega \mathcal{M}^{(k)\dagger} = \mathcal{V}^{(k) } \mathcal{T}^{(k)} \mathcal{V}^{(k) \dagger}
\label{meig}
\end{equation}
with $\mathcal{V}^{(k)}$ as the local PMs in the $k$th section, and where
\begin{equation}
 \mathcal{T}^{(k)} = \mathrm{diag}\left[\tau^{(k)}_{ 1}, \tau^{(k)}_{ 2}, \cdots, \tau^{(k)}_{ D}\right] 
 \label{meigt}
\end{equation} 
is a diagonal matrix of their GDs in the $k$th section.  
With $\sum_{i} \tau^{(k)}_{i} = 0$, we have $\mathrm{tr} \left( \mathcal{T}^{(k)} \right) = 0 $ and 
\begin{equation}
\mathrm{tr}\left( j \mathcal{M}^{(k)}_\omega \mathcal{M}^{(k)\dagger} \right) = 0.
\end{equation}
Physically, the $i$th local PM experiences an uncoupled GD $\tau^{(k)}_{i}$ without mixing with other modes. 
Because the diagonal matrices $ \mathcal{T}^{(k)}$ are real matrices, all matrices  $j \mathcal{M}^{(k)}_\omega \mathcal{M}^{(k)\dagger}$, $k = 1, \dots, K$, are Hermitian. 

When $K$ sections of MMF are cascaded together, the overall  propagation matrix becomes
\begin{equation}
\mathcal{M}^{(t)} = \mathcal{M}^{(K)} \mathcal{M}^{(K-1)} \cdots \mathcal{M}^{(2)}  \mathcal{M}^{(1)}.
\end{equation}
The overall PMs and their GDs correspond to the eigenvectors and eigenvalues of \cite{fan05, shemirani09}
\begin{equation}
\mathcal{G} = j \mathcal{M}^{(t)}_\omega \mathcal{M}^{(t)\dagger}.
\end{equation}
Because
\begin{multline}
\mathcal{M}^{(t)}_\omega = \mathcal{M}^{(K)}_\omega \mathcal{M}^{(K-1)} \cdots \mathcal{M}^{(2)}  \mathcal{M}^{(1)} \\
                         + \mathcal{M}^{(K)} \mathcal{M}^{(K-1)}_\omega   \cdots \mathcal{M}^{(2)}\mathcal{M}^{(1)} +  \\ 
                \cdots + \mathcal{M}^{(K)} \mathcal{M}^{(K-1)} \cdots \mathcal{M}^{(2)}  \mathcal{M}^{(1)}_\omega,
\end{multline}
we obtain
\begin{multline}
\mathcal{G}   =  j \mathcal{M}^{(K)}_\omega \mathcal{M}^{(K)\dagger} + j \mathcal{M}^{(K)} \mathcal{M}^{(K-1)}_\omega \mathcal{M}^{(K-1)\dagger}  \mathcal{M}^{(K)\dagger} +  \\
  \quad \cdots + j \mathcal{M}^{(K)} \mathcal{M}^{(K-1)} \cdots \mathcal{M}^{(2)}  \mathcal{M}^{(1)}_\omega  \mathcal{M}^{(t)\dagger} . 
 \label{mexpand}
\end{multline}

From \eqn{mexpand}, the overall matrix $j \mathcal{M}^{(t)}_\omega \mathcal{M}^{(t)\dagger}$ is the summation of $K$ random matrices.
All those $K$ matrices have eigenvalues identical to those of \eqn{meig} and \eqn{meigt}.
However, their eigenvectors are independent of each other.
The first matrix $j \mathcal{M}^{(K)}_\omega \mathcal{M}^{(K)\dagger}$ has eigenvectors derived from  $\mathcal{V}^{(K)}$.
The second matrix  $j \mathcal{M}^{(K)} \mathcal{M}^{(K-1)}_\omega \mathcal{M}^{(K-1)\dagger}  \mathcal{M}^{(K)\dagger}$ has eigenvectors derived from $\mathcal{M}^{(K)}  \mathcal{V}^{(K-1)}$.
Matrices  $\mathcal{V}^{(K)}$ and $\mathcal{M}^{(K)}  \mathcal{V}^{(K-1)}$ are both unitary matrices and are obviously independent of each other.
All the $K$ matrices summed to form $\mathcal{G}$ are independent of each other with eigenvalues given by the vectors $\vec{\tau}^{(k)}$, $k = 1, \dots, K$. 
Even for the case that all vectors $\vec{\tau}^{(k)}$ are deterministic and identical, all the $K$ matrices summed to form $\mathcal{G}$ are independent, owing to the different directions of their independent eigenvectors.

The matrix elements of $\mathcal{G}$ should be identically distributed Gaussian random variables from the central limit theorem (CLT).
The matrix elements of $\mathcal{G}$, $g_{i, j}$, $i, j = 1, \dots, D$, are the summation of $K$ identically distributed random variables, as seen from \eqn{mexpand}.
If $K$ is very large, $g_{i, j}$ are Gaussian random variables from the CLT.
Because all $K$ component matrices in \eqn{mexpand} are Hermitian, $\mathcal{G}$ is a Hermitian matrix.
The diagonal elements $g_{i,i}$, $i = 1, \dots, D$, are all real Gaussian random variables with variance equal to $\sigma_g^2$.
All non-diagonal elements $g_{i, j}$, $i \neq j$, are complex Gaussian random variables with independent real and imaginary parts, which have variance equal to $\sigma_g^2/2$. 
Thus, the elements $g_{i, j}$, $i \neq j$ has variance $\sigma_g^2$.
The value  $\sigma_g^2$ depends on the number of modes $D$, the number of sections $K$, and the variances of the uncoupled GDs described by  $\vec{\tau}^{(k)}$.
\revisenew{If the $D$ vectors in $\mathcal{V}^{(k)}$ given by \eqn{meig} are assumed to be independent of each other, it can be shown that}
\begin{equation}
\sigma_g^2 = \frac{1}{D^2} \sum_{k=1}^K || \vec{\tau}^{(k)} ||^2 = \frac{1}{D}\sum_{k = 1}^K \sigma^2_{\tau^{(k)}},
\label{eq:sigmag}
\end{equation}
where $\sigma^2_{\tau^{(k)}}$, $k = 1, \dots, K$, are the variances of the GDs in the sections.
If all $K$ sections have the same modal GD profiles, we have 
\begin{equation}
\sigma_g^2 = \frac{K}{D} \sigma^2_{\tau},
\label{eq:sigmag1}
\end{equation}
 where $\sigma^2_{\tau}$ are the GD variances in all sections.

In random matrix theory, the matrix $\mathcal{G}$ is described as a Gaussian unitary ensemble \cite[Sec. 2.5]{mehta}.
Typically, a Gaussian unitary ensemble does not have any constraint aside from the variance of its Gaussian elements.
However, in \eqn{mexpand}, the matrix components have zero trace so that
\begin{equation}
\mathrm{tr} \left( \mathcal{G} \right) = 0.
\label{trG}
\end{equation}
In other words, $\mathcal{G}$ is a zero-trace Gaussian unitary ensemble.
The GDs in a MMF are statistically described by the eigenvalues of the zero-trace Gaussian unitary ensemble. 

\revisenew{
The assumption that the diagonal and off-diagonal elements of $\mathcal{G}$ have the same variance $\sigma_g^2$ \eqn{eq:sigmag1} is valid only if the $D$ orthogonal vectors in $\mathcal{V}^{(k)}$ are independent of each other. 
However, the  condition of orthogonality implies that the $D$th vector is determined by the other $D-1$ vectors.
Using numerical simulation, we have found that all diagonal elements of $\mathcal{G}$ have equal variance of $(1 - D^{-1})\sigma_g^2$, all off-diagonal elements of $\mathcal{G}$ have equal variance of $[1 + 1/D(D-1)]\sigma_g^2$, and the average variance of all elements of $\mathcal{G}$ is $\sigma_g^2$ \eqn{eq:sigmag1}.
 The theory of \cite[Sec. 14.3]{mehta} is able to describe  a random Hermitian matrix in which different elements have different variances.
  Numerical simulations of the zero-trace Gaussian unitary ensemble with unequal variances that is considered here show no observable differences from analytical results derived assuming all matrix elements have equal variance \cite[Sec. 3.3]{mehta}.
}

\section{Modal Dispersion in Few-Mode Fibers}
\label{sec:fmf}

In the regime of strong mode coupling, the PMs and their GDs are given by the eigenvectors and eigenvalues of the zero-trace Gaussian unitary ensemble described by \eqn{mexpand} and \eqn{trG}.
Without loss of generality, after normalization, the elements of $\mathcal{G}$ may be assumed to be zero-mean identically distributed Gaussian random variables with variance $\sigma_g^2 = 1/2$, similar to the classic normalization of Mehta \cite{mehta}\footnote{In some mathematical literature, which is relevant to Gaussian orthogonal ensembles
but not Gaussian unitary ensembles, the matrix elements are assumed to
have unit variance. Proportionality constants are ignored in some of those references.}.
Before normalization, $\sigma_g^2$ is given by either \eqn{eq:sigmag} or \eqn{eq:sigmag1}.
The diagonal elements of $\mathcal{G}$ are real with a variance of $\sigma_g^2 = {1}/{2}$.
The off-diagonal elements of $\mathcal{G}$ are complex Gaussian distributed with independent real and imaginary parts, each having a variance of ${1}/{4}$.
From \eqn{eq:sigmag1}, each section can be taken to have a normalized standard deviation of GD given by \revise{2.3}{$\sigma_\tau = \sqrt{D/2K}$}. 
With this normalization, the notation  in this section is similar to that in Mehta \cite{mehta}.

\subsection{Joint Probability Density}

The joint p.d.f.~for a Gaussian unitary ensemble without the zero-trace constraint is well-known. 
The ordered joint p.d.f.~of the eigenvalues a $D \times D$ Gaussian unitary ensemble is  \cite[Sec. 3.3]{mehta} \cite{ginibre65}
\begin{equation}
\alpha_D \prod_{D \geq i > j > 0} (\lambda_i - \lambda_j)^2 \exp\left( - \sum_{i = 1}^D \lambda_i^2 \right),
\label{jpdf_everytr}
\end{equation}
where the eigenvalues \revise{2.4}{possess the order constraint} $\lambda_1 \leq \lambda_2 \leq \cdots \leq \lambda_D$ and are all real valued, and $\alpha_D$ is a constant such that the joint p.d.f.~integrates to unity.
The eigenvalues are the normalized GDs. 
An analytical expression for $\alpha_D$ can be found in \cite[Theorem 3.3.1]{mehta}. 
Because a permutation with a different ordering of eigenvalues is equivalent to any other permutation,
 the unordered joint p.d.f.~is just $1/D!$ of \eqn{jpdf_everytr} but without the order constraint \cite[ch. 5]{mehta}. 

With zero trace, 
\begin{equation}
\mathrm{tr}(\mathcal{G}) = \lambda_1 + \lambda_2 + \cdots + \lambda_{D} = 0,
\label{zerotr_lambda}
\end{equation}
 the ordered joint p.d.f.~of $\lambda_1, \dots, \lambda_{D-1}$ becomes
\begin{equation}
p_D(\lambda_1,\dots,\lambda_{D-1}) = \beta_D\!\!\prod_{D \geq i > j > 0}\!\!(\lambda_i - \lambda_j)^2 \exp\left( - \sum_{i = 1}^D \lambda_i^2 \right),
\label{jpdf_zerotr}
\end{equation}
%
%
%
\remark{Delete for 2.5.}
with the order constraint 
\begin{equation}
\lambda_1 \leq \lambda_2 \leq \cdots \leq \lambda_D,
\label{lambda_order}
\end{equation}
where the constant $\beta_D$ is determined by requiring  \eqn{jpdf_zerotr} to integrate to unity, but is not the same as the $\alpha_D$ in \eqn{jpdf_everytr}.
The unordered joint p.d.f.~is the same as \eqn{jpdf_zerotr} but is a factor of $1/D!$  smaller and does not have the order constraint \eqn{lambda_order}.  
The statistical properties of GD are fully specified  by the joint p.d.f.~\eqn{jpdf_zerotr} with the constraint \eqn{zerotr_lambda}.

\remark{Delete for 2.6.}

\subsection{Two-Mode Fiber}

Two-mode fiber is the simplest case, and may correspond to the two polarization modes in a SMF, i.e., the well-known PMD problem. 
The purpose here is not to derive new properties of PMD, but to verify that the general random matrix model is applicable to PMD.

With $\lambda_2 = -\lambda_1$, the p.d.f.~\eqn{jpdf_zerotr} for $D = 2$ becomes
\begin{equation}
p_2(\lambda_1) = \beta_2 4 \lambda_1^2 e^{-2\lambda_1^2}.
\end{equation}
As in the PMD literature, we define $\lambda_{1,2} = \pm \frac{1}{2} \tau$ and find $\beta_2 = \sqrt{2/\pi}$, obtaining
\begin{equation}
p_2(\tau) = \sqrt{\frac{2}{\pi}} \tau^2 e^{-\tau^2/2},  \tau \geq 0,
\label{eq:p2}
\end{equation}
which is the well-known Maxwellian distribution with normalized mean DGD of $\bar{\tau} = 2 \sqrt{2/\pi} = 1.60$.
Random matrix models specialized to the two-mode case were used to derive the Maxwellian distribution in \cite{karlsson01,poole97}.
The second moment of both $\lambda_1$ and $\lambda_2$ is $3/4$. 

\subsection{Three-Mode Fiber}

Three-mode fiber is the next-simplest case, with $\lambda_3 = -(\lambda_1 + \lambda_2)$. 
The joint p.d.f.~\eqn{jpdf_zerotr} for $D = 3$ of \eqn{lambda_order} is
\begin{align}
p_3(\lambda_1, \lambda_2) =  & \beta_3 (\lambda_2 - \lambda_1)^2(2\lambda_1 + \lambda_2)^2(\lambda_1 + 2\lambda_2)^2  \nonumber\\
  & \qquad \times 
e^{-\lambda_1^2 - \lambda_2^2 - (\lambda_1 + \lambda_2)^2}.
\end{align}

Without the order constraint \eqn{lambda_order}, the marginal p.d.f.~of the GDs is
\begin{equation}
p_{3 \lambda}(\lambda) = \frac{1}{6}\int_{-\infty}^{+\infty} p_3(\lambda, \lambda_2) \ud \lambda_2.
\end{equation}
Some algebra\footnote{\revise{2.8}{
All the calculations performed for $D \geq 3$ require three steps. 
In the first step, $\lambda_1^2 + \lambda_2^2 + \cdots + \lambda_{D-1}^2 + (\lambda_1 + \lambda_2 + \cdots + \lambda_{D-1})^2$ is linearly transformed to $D/(D-1) \times \lambda_1^2 + x_2^2 + \cdots + x_{D-1}^2$. 
In the second step, the linear transform is substituted into $\prod_{D \geq i > j > 0} (\lambda_i - \lambda_j)^2$ , which may be expanded to a summation of terms in the form of  $c_{f_1, f_2, .., f_{D-1}}\lambda_1^{f_1}x_2^{f_2} \cdots x_{D-1}^{f_{D-1}}$, where $f_i$ are the exponents  and $c_{f_1, f_2, .., f_{D-1}}$ are the corresponding coefficients. 
The last step is the integration over $x_2, x_3, \dots, x_{D-1}$ using \cite[Sec. 3.461]{table}.
These calculations are tedious, but can be performed using symbolic mathematical software, such as Maple or MuPAD.
}} 
yields the constant $\beta_3 = 4\sqrt{3}/\pi$ and
\begin{equation}
p_{3 \lambda}(\lambda) = \frac{1}{16}\sqrt{\frac{6}{\pi}} \left( 27 \lambda^4 - 18\lambda^2 + 5 \right) \exp \left( -\frac{3}{2} \lambda^2   \right).
\label{p3l}
\end{equation}

\begin{figure}[t]
\centering{
 \includegraphics[width = 0.45 \textwidth]{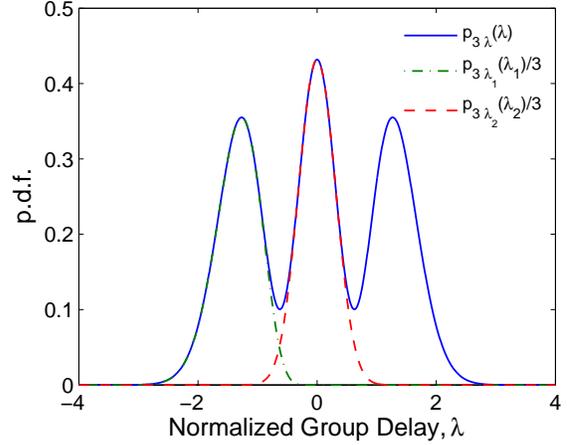}}
 \caption{Statistics of the normalized GDs for three-mode fiber, including the marginal p.d.f.~$p_{3 \lambda}(\lambda)$, the p.d.f.~$p_{3 \lambda_1}(\lambda_1)$ of the smallest delay, and the p.d.f.~$p_{3 \lambda_2}(\lambda_2)$  of the middle delay.
 To facilitate comparison, both $p_{3 \lambda_1}(\lambda_1)$ and $p_{3 \lambda_2}(\lambda_2)$ are scaled by a factor of ${1}/{3}$.
 }
\label{fig:p3}
\end{figure}

\Fig{fig:p3} plots the marginal p.d.f.~$p_{3 \lambda}(\lambda)$, which exhibits three peaks, corresponding to the values where $\lambda_1$, $\lambda_2$, $\lambda_3$ are concentrated.
The p.d.f.~$p_{3 \lambda}(\lambda)$ is symmetrical with respect to $\lambda = 0$ due to the symmetric nature of the three eigenvalues for the $3 \times 3$ random matrix $\mathcal{G}$.
The middle eigenvalue $\lambda_2$ is concentrated near zero.
The variance of $\lambda$ is $\sigma_{\lambda}^2 = 4/3$.

The p.d.f.~of the eigenvalue $\lambda_1$, corresponding to the smallest delay, can also be found using \eqn{jpdf_zerotr} with the order constraint \eqn{lambda_order}. 
The condition $\lambda_1 \leq 0$ is required in order to conform to the zero-trace constraint \eqn{zerotr_lambda}. 
The p.d.f.~of the smallest eigenvalue is
\begin{equation}
p_{3 \lambda_1} (\lambda_1) = \int_{\lambda_1}^{-\frac{\lambda_1}{2}} p_3(\lambda_1, \lambda_2) \ud \lambda_2,
\end{equation}
or
\begin{multline}
p_{3 \lambda_1} (\lambda_1) =
3 p_{3 \lambda} (\lambda_1)
\mathrm{erf}\left(\frac{3 |\lambda_1|}{\sqrt{2}} \right) \\
    - \frac{9 \sqrt{3} \lambda_1(3\lambda_1^2-5)}{ 8 \pi} \exp(-6 \lambda_1^2), \\ \qquad \lambda_1  \leq 0. \qquad
     \label{p3l1}
\end{multline}

Due to symmetric nature of $\lambda_1$ and $\lambda_3$, we have $p_{3 \lambda_3} (\lambda_3) = p_{3 \lambda_1} (-\lambda_3)$.
\Fig{fig:p3} also shows $\frac{1}{3} p_{3 \lambda_1} (\lambda_1)$ where the p.d.f.~ is scaled by a factor ${1}/{3}$ such that $p_{3 \lambda_1} (\lambda_1)$ given by \eqn{p3l1} is nearly the same as $p_{3 \lambda}(\lambda)$ given by \eqn{p3l} near the first peak of  $p_{3 \lambda}(\lambda)$.

Similarly, the p.d.f.~of the middle eigenvalue $\lambda_2$ can be found by
\begin{equation}
p_{3 \lambda_2} (\lambda_2) = \int_{-\infty}^{-|\lambda_2|} p_3(\lambda_1, \lambda_2) \ud \lambda_1,
\end{equation}
or
\begin{multline}
p_{3 \lambda_2} (\lambda_2) = 
3 p_{3 \lambda} (\lambda_2)
 \mathrm{erfc}\left(\frac{3 |\lambda_2|}{\sqrt{2}} \right) \\
     - \frac{9 \sqrt{3} |\lambda_2|(3\lambda_2^2-5)}{ 8 \pi} \exp(-6 \lambda_2^2), 
     \label{p3l2}
\end{multline}

Comparing \eqn{p3l}, \eqn{p3l1}, and \eqn{p3l2}, the marginal p.d.f.~$p_{3 \lambda} (\lambda)$ \eqn{p3l} is found to be the combination of $p_{3 \lambda_1} (\lambda_1)$,  $p_{3 \lambda_2} (\lambda_2) $, and  $p_{3 \lambda_3} (\lambda_3)$:

\begin{equation}
p_{3 \lambda} (\lambda) = \frac{1}{3} \left[     
             p_{3 \lambda_1} (\lambda) + p_{3 \lambda_2} (\lambda)  + p_{3 \lambda_3} (\lambda)
\right].
\end{equation}

\Fig{fig:p3} also shows $\frac{1}{3} p_{3 \lambda_1} (\lambda_1)$ and $\frac{1}{3} p_{3 \lambda_2} (\lambda_2)$ which are concentrated near the corresponding peaks of the marginal p.d.f.~$ p_{3 \lambda} (\lambda)$, confirming that each peak of the marginal p.d.f.~$ p_{3 \lambda} (\lambda)$ corresponds to an individual eigenvalue.
The statistical parameters of the normalized GDs of a three-mode fiber are presented in Table \ref{tab:3mode}.
In \fig{fig:p3}, the peak for the middle delay $\lambda_2$ is narrower than those for the maximum and minimum delays $\lambda_1$ and $\lambda_3$.
In Table \ref{tab:3mode}, the variance of $\lambda_2$ is smaller than the variance of $\lambda_1$ or $\lambda_3$.

\begin{table}
\centering{
\caption{Statistical parameters for the normalized GDs of a three-mode fiber}
\label{tab:3mode}
\begin{tabular}{lll}
\hline 
p.d.f.  &   Mean    & Variance \\
\hline 
$p_{3 \lambda}(\lambda)   $  & $0 $        & $\frac{4}{3}$ \\
$p_{3 \lambda_1}(\lambda_1)$ & $-\frac{27}{16}\sqrt{\frac{2}{\pi}}$  & $\frac{4}{3} + \frac{9\sqrt{3}}{8\pi} - \frac{729}{128\pi} = 0.1407 $  \\
$p_{3 \lambda_2}(\lambda_2) $&  $0 $      & $\frac{4}{3} - \frac{9\sqrt{3}}{4 \pi} = 0.0928$ \\
$p_{3 \lambda_3}(\lambda_3) $&  $\frac{27}{16}\sqrt{\frac{2}{\pi}}$  & $0.1407 $  \\
$p_{3 (\lambda_3 -\lambda_1)}(\lambda) $& $\frac{27}{8}\sqrt{\frac{2}{\pi}} $ & $4 - \frac{729}{32 \pi} + \frac{27\sqrt{3}}{4 \pi} = 0.4700 $\\
\hline
\end{tabular}
}
\end{table}

The difference between the maximum and minimum eigenvalues is the normalized delay spread of the MMF. 
The p.d.f.~of the delay spread for three-mode MMF is
\begin{equation}
p_{3 (\lambda_3 -\lambda_1)}(\lambda) = \frac{1}{2} \int_{-\lambda/3}^{\lambda/3} p_3\left(\frac{\lambda-\lambda_2}{2}, \lambda_2 \right) \ud \lambda_2
\end{equation}
or
\begin{multline}
p_{3 (\lambda_3 -\lambda_1)}(\lambda) = 
   \frac{\sqrt{3}}{4\pi} (\lambda^5 - 9 \lambda^3) \exp\left(-\frac{2}{3} \lambda^2 \right)
     \\
      + \frac{1}{8} \sqrt{\frac{2}{\pi}} 
     \left( \lambda^6 -6\lambda^4 + 27 \lambda^2 \right)     
     \exp\left(-\frac{1}{2} \lambda^2 \right) \mathrm{erf} \left(\frac{\lambda}{\sqrt{6}}\right), \\ \quad \lambda > 0.
\end{multline}
The statistical parameters of the delay-spread are also given in Table \ref{tab:3mode}.
For three-mode fiber, many properties of the eigenvalues or normalized GDs can be computed  analytically in closed form.

\subsection{Four-Mode Fiber}

Four-mode fiber is a particularly simple case beyond the two-mode fiber, as it represents a fiber with two spatial modes and two polarizations. 
Fibers with two spatial modes have been used
for dispersion compensation \cite{poole92, huttuen05} and for transmission experiments \cite{li11, salsi11}. 
We note, however, that a weakly guiding fiber with circular core cannot support exactly two spatial modes \cite{gloge70}.
Although the relatively short distances used in \cite{li11, salsi11} may not be sufficient to ensure strong mode coupling, the strong-coupling regime can be expected in  future long-distance transmission systems. 

The p.d.f.~\eqn{jpdf_zerotr} for $D = 4$ becomes
\begin{equation}
p_4(\lambda_1, \lambda_2, \lambda_3) = \beta_4 \prod_{4 \geq i, j \geq 1}(\lambda_i - \lambda_j)^2
e^{-\lambda_1^2 - \lambda_2^2 - \lambda_3^2 - \lambda_4^2},
\label{eq:p4_1234}
\end{equation}
with the zero trace constraint  $\lambda_4 = -\lambda_1 - \lambda_2 - \lambda_3$. 

Using the unordered joint p.d.f., the marginal p.d.f.~of $\lambda$ is
\begin{equation}
p_{4 \lambda}(\lambda) = \frac{1}{4!} \int_{-\infty}^{+\infty} \int_{-\infty}^{+\infty} p_4(\lambda, \lambda_2, \lambda_3) \ud \lambda_2 \ud \lambda_3.
\end{equation}
After some calculations, we obtain
\begin{equation}
p_{4 \lambda}(\lambda) = {\frac {2 \sqrt {3}{e^{-4/3{\lambda}^{2}}}}{ \sqrt {\pi }}}\!\!\left( \frac{4096}{6561}{\lambda}^{6}-\frac{1024}{729}{\lambda}^{4}+\frac{80}{81}{\lambda}^{2}+\frac{5}{81} \right),
\label{eq:p4l}
\end{equation}
which has a variance of  $15/8$. 

\begin{figure}[t]
\centering{
 \includegraphics[width = 0.45 \textwidth]{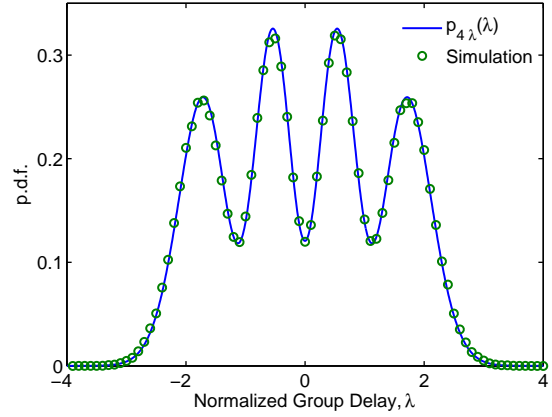}}
 \caption{The marginal p.d.f.~of the normalized GDs for a four-mode fiber, comparing analysis and simulation.}
\label{fig:p4}
\end{figure}

\Fig{fig:p4} shows the marginal p.d.f.~of the  normalized GDs in a four-mode fiber given by  \eqn{eq:p4l}.
The marginal p.d.f.~has four peaks, corresponding to the GD of four different PMs.

To verify the marginal p.d.f.~in \fig{fig:p4}, the modal dispersion of a four-mode MMF has been simulated.
The fiber has $K = 256$ independent sections. 
In each section, the four modes are chosen to have deterministic delays of $+\tau, +\tau, -\tau$, and $-\tau$ where $\tau = \sqrt{2/K}$ to ensure that the elements of $\mathcal{G}$ have a variance of $\sigma_g^2 = {1}/{2}$.
This particular choice could describe a fiber where, in each section, the DGD between the two polarization modes is negligible compared to that between the two spatial modes. 
The random unitary matrices $\mathcal{U}^{(k)}$ and  $\mathcal{V}^{(k)}$, $k = 1, \dots, K$, are first initialized by $4 \times 4$ random complex Gaussian matrices and then converted to unitary matrices using the Gram-Schmidt process \cite[Sec. 5.2.8]{golub3}. 
All sections have independent matrices $\mathcal{U}^{(k)}$ and  $\mathcal{V}^{(k)}$.
A total of $400,000$ eigenvalues are used in the curve shown in Fig. 2. 

In \fig{fig:p4}, the simulation results show excellent agreement with the analytical p.d.f.~\eqn{eq:p4l}.
Although the modes in each section have only two GDs, with strong mode coupling, a p.d.f.~having four peaks is obtained. 
In the strong-coupling regime, similar results would be obtained using any uncoupled GDs in each section, provided that the four GDs sum to zero and have a variance of $2/K$.
For example, the four modes may have modal delays of $+\tau_1, +\tau_1, +\tau_1$, and $-3 \tau_1$ with $\tau_1 = \sqrt{2/3K}$.
Similar results may also be obtained if the GDs in each section are, for example, $+\tau_2$, $+\tau_2$, $-\tau_2$, $-\tau_2$, where $\tau_2$ follows a statistical distribution with second moment $2/K$.

The marginal p.d.f.~of the smallest or largest eigenvalues $\lambda_1$ and $\lambda_4$ may be found by suitable integration of \eqn{eq:p4_1234}.
Unlike the case of three-mode fiber, it does not seem possible to obtain closed-form expressions for the individual marginal p.d.f.'s of the ordered eigenvalues.

\subsection{Other Few-Mode Fibers}

The marginal p.d.f.'s~of the GDs in MMF with larger number of modes may also be obtained analytically.
Following the above procedure, the marginal p.d.f.'s~of fiber with five, six, and seven modes are
\begin{multline}
p_{5\lambda}(\lambda) = \frac{\sqrt {5} {e^{-5/4\,{\lambda}^{2}}}}{\sqrt {\pi }}\,\left(\frac{78125}{196608}\,{\lambda}^{8} -\frac{15625}{8192}\,{\lambda}^{6}\right.\\
\left.+\frac{24375}{8192}\,{\lambda}^{4}-\frac{1975}{2048}\,{\lambda}^{2}+\frac{903}{4096} \right),
\end{multline}
\begin{multline}
p_{6\lambda}(\lambda) = {\frac {\sqrt {30}{{e^{-6/5{\lambda}^{2}}}}}{\sqrt {\pi }}}\left( \frac{13436928}{244140625}{\lambda}^{10}-\frac{4478976}{9765625}{\lambda}^{8} \right. \\ \left. +  \frac{2581632}{1953125}{\lambda}^{6}-\frac{102816}{78125}{\lambda}^{4}+\frac{7812}{15625}{\lambda}^{2}+\frac{644}{15625}\right),
\end{multline}
and
\begin{multline}
p_{7\lambda}(\lambda) = {\frac {\sqrt {42} {e^{-7/6{\lambda}^{2}}}}{\sqrt{\pi}}}\left(  \frac{1977326743}{146932807680}\,{\lambda}^{12}\right. \\ -\frac{282475249}{1632586752}\,{\lambda}^{10}+\frac{98001617}{120932352}\,{\lambda}^{8} -\frac{47883143}{30233088}\,{\lambda}^{6}  \\ \left. +\frac{17707375}{13436928}\,{\lambda}^{4}-\frac{212219}{746496}\,{\lambda}^{2}+\frac{88175}{1492992}\right),
\end{multline}
respectively.
The variances of these distributions are $12/5$, $35/12$, and $24/7$, respectively.
For all cases of $D$ from two to seven modes, the variances are given by $\frac{1}{2}(D - D^{-1})$, \revisenew{a reduction by a factor of $1 - D^{-2}$ compared with the case without the zero-trace constraint.}

\begin{figure}[t]
\centering{
 \includegraphics[width = 0.45 \textwidth]{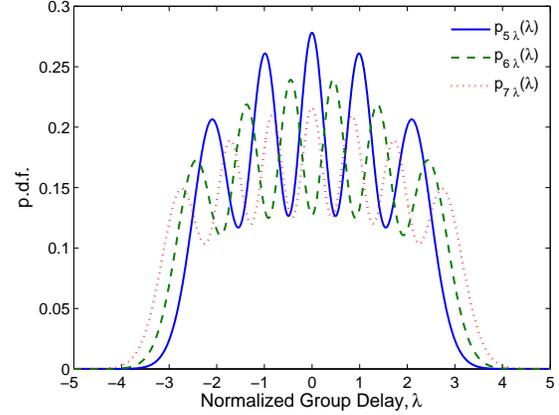}}
 \caption{The marginal p.d.f.~of normalized GDs for fibers with five, six, and seven modes. }
\label{fig:p567}
\end{figure}

\Fig{fig:p567} shows the marginal p.d.f.~of the normalized GDs of fibers with five, six, and seven modes. 
The number of peaks in the marginal p.d.f.~is the same as the number of modes.
In general, the peaks closer to $\lambda = 0$ are both higher and narrower than those farther from the origin.
Those peaks cause ripples to appear in the marginal p.d.f., and the ripples are still significant in seven-mode MMF.
As the number of modes increases, the peaks in the marginal p.d.f.~move closer together and merge.
As the peaks become indistinguishable for fibers having many modes, the marginal p.d.f.~should approach a limiting distribution.


\revisenew{
Numerical simulations, similar to those in \fig{fig:p4}, have been used to verify the analytical p.d.f.'s~of \fig{fig:p567}.  
In all the cases, simulation and theory match with each other.

Numerical simulations have been conducted to further verify the variance reduction factor of $1 - D^{-2}$, which is most significant for $D \leq 3$. 
Random realizations of zero-trace matrices $\mathcal{G}$ of the form \eqn{mexpand} exhibit no observable variance reduction, although the empirically estimated p.d.f.'s, when scaled by this reduction factor, are found to match with \eqn{eq:p2} and \eqn{p3l}.

The reduction of variance may be seen as related to degrees of freedom. 
A random Hermitian matrix without zero-trace constraint has $D^2$ degrees of freedom, corresponding to $D(D-1)/2$ complex off-diagonal elements and $D$ real diagonal elements. 
The zero-trace constraint reduces the degrees of freedom by one, proportionally affecting a fraction $1/D^2$ of matrix elements. 
The zero-trace constraint from \eqn{jpdf_everytr} to \eqn{jpdf_zerotr} reduces $D$ degrees of freedom to $D-1$ degrees of freedom, proportionally affecting a fraction $1/D$ of  matrix elements.
While analytical results scaled to the same variance are consistent with numerical simulations, the variance reduction factor of $1-D^{-2}$ requires further study.

In the simplest case of $D = 2$, a zero-trace Gaussian unitary ensemble can be generated numerically  by three methods.
The first method is based on  $\mathcal{G}$ given by the summation \eqn{mexpand}, for example, with $\sigma_g^2 = 1/2$.
The second method is based on a random $2 \times 2$ Hermitian matrix  $\mathcal{A}$ but with $a_{2,2}$ replaced by $-a_{1,1}$.
The third method is based  on generating random $2 \times 2$ Hermitian matrix  $\mathcal{A}$, finding its eigenvalues $\lambda_1$ and $\lambda_2$, and selecting those with $\left|\lambda_1 + \lambda_2\right|$ smaller than a small number. 
In the second and third methods, the elements of the matrix $\mathcal{A}$ have variance of $1/2$.
The first two methods give eigenvalues with the same variance but the third method gives eigenvalues with a variance $3/4$ time smaller than the first two methods. 
In the third method, the variance of the diagonal elements are $1/4$ and the off-diagonal elements have variance of $1/2$. 
The zero-trace constraint on the eigenvalues \eqn{zerotr_lambda} reduces the variance of the matrix elements but the p.d.f.~maintains the same shape. 

The variance reduction of the diagonal elements is due to the zero-trace constraint \eqn{zerotr_lambda}, which selects those matrices with smaller diagonal elements (in the general case, a factor of $1 - D^{-1}$ smaller). 
The average variance among all elements is a factor $1 - D^{-2}$ smaller, the same as the reduction factor for the variance of the eigenvalues. 

As the average variance for all matrix elements of $\mathcal{G}$ is $\sigma_g^2$ given by \eqn{eq:sigmag1}, 
for all MMF studies in this section, and without normalization}, the variance of the GD is
\begin{equation}
\sigma_\mathrm{gd}^2 =  K \sigma_\tau^2
\label{eq:sigmamd}
\end{equation}
with $\sigma^2_\tau$  defined by \eqn{eq:sigmag1}.
If the GD is characterized by its standard deviation $\sigma_\mathrm{gd}$, it is always proportional to the square root of the number of independent MMF sections.

\section{Modal Dispersion in Many-Mode Fibers}
\label{sec:infmode}

With a large number of modes, a Gaussian unitary ensemble without the zero-trace constraint is described by a semicircle distribution with radius $\sqrt{2 D}$ \cite[Sec. 4.2]{mehta}.
With the normalization used in Sec. \ref{sec:fmf}, the variance of the eigenvalues is $D/2$.
This semicircle law was first derived by Wigner for large random matrices \cite{wigner55, wigner58}.
The Wigner semicircle law is universally valid for many different types of large random matrices \cite{erdos10, tao09}.
A Gaussian unitary ensemble, even with the zero-trace constraint \eqn{zerotr_lambda}, should follow the semicircle distribution. 
As an alternative to considering $\mathcal{G}$ as a Gaussian unitary ensemble, a more straightforward derivation using the CLT for free random variables is given in a later part of this section.

In free probability theory, free random variables are equivalent to statistically independent large random matrices \cite{voiculescu92,nica06}.
The CLT for the summation of free random variables gives the semicircle distribution \cite{voiculescu91,voiculescu92}.
The matrix $\mathcal{G}$ \eqn{mexpand} is the summation of many independent random matrices.
The CLT for free random variables states the following: Let $\mathcal{X}_k$, $k = 1, \dots, K$, be identically distributed independent zero-mean free random variables with unit variance. 
The summation 
\begin{equation}
\mathcal{Y}_K = \frac{\mathcal{X}_1 + \mathcal{X}_2 + \cdots + \mathcal{X}_K}{\sqrt{K}}
\label{freeY}
\end{equation} 
is described by semicircle distribution with radius of two and unit variance
\begin{equation}
p_Y(r) = \left\{ \begin{array}{ll}
                  \frac{1}{2 \pi} \sqrt{4 - r^2}  & |r| < 2 \\
                  0    & \mathrm{otherwise}
		   \end{array} \right.
\end{equation}
as $K$ approaches infinity. 

In the above theorem, when free random variables are represented by large random matrices, the distribution of a free random variable is equivalent to the distribution for the eigenvalues of the random matrices.
When the CLT of free random variables is applied to $\mathcal{G}$ given by \eqn{mexpand}, if the variance of the zero-mean GD per section is $\sigma^2_{\tau}$ for all $K$ sections, the eigenvalues of $\mathcal{G}$ are described by a semicircle distribution with radius  $2 \sqrt{K} \sigma_\tau$ and variance  $K \sigma_\tau^2$.
Equivalently, the GD of the MMF has a semicircle distribution with variance $K \sigma_\tau^2$.
Note that the normalization used in this section based on the eigenvalues of  $\mathcal{X}_k$ and $\mathcal{Y}_K$ in \eqn{freeY} is customary in free probability theory.
However,  the normalization used in Sec.~\ref{sec:fmf} is based on the matrix elements of $\mathcal{G}$, similar to that in Mehta \cite{mehta}.

\begin{figure}[t]
\centering{
 \includegraphics[width = 0.45 \textwidth]{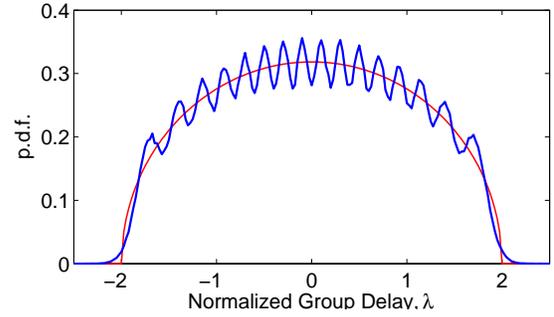} \\
 (a) $D = 16$ \\
 \includegraphics[width = 0.45 \textwidth]{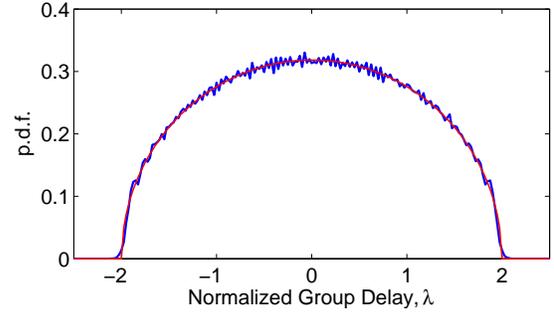} \\
 (b) $D = 64$ \\
  \includegraphics[width = 0.45 \textwidth]{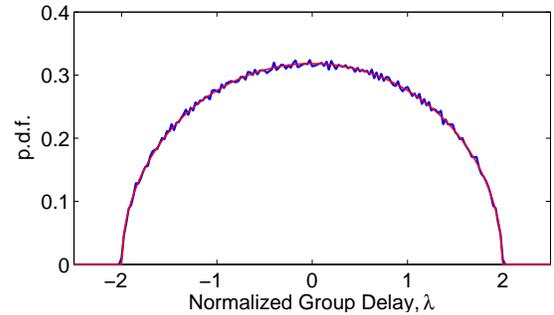} \\
 (c) $D = 512$}
 \caption{Simulated marginal p.d.f.~of normalized GDs (ripply blue curves) compared with semicircle distribution (smooth red curves) in MMF with (a) 16, (b) 64, and (c) 512 modes. 
 }
\label{fig:p64}
\end{figure}

\Fig{fig:p64} compares the simulated marginal p.d.f.~of MMFs having $D=16$, $64$, and $512$ modes to the  semicircle distribution.
Each MMF is comprised of  $K=256$ sections.
In each section, the GDs are deterministic, with the first $D/2$ modes with a delay of $\tau$ and the other $D/2$ modes with a delay of $-\tau$.
The simulated curves are obtained from $1,600,000$ and $640,000$ and $102,400$ eigenvalues for $D = 16$ and $64$ and $512$, respectively, using a step size of $0.025$ along the $\lambda$ axis.
The normalization $\tau = 1/\sqrt{K}$ is made to facilitate comparison with a semicircle distribution with radius of $2$.
The model here is valid as long as the component matrices in \eqn{mexpand} may be modeled as free random variables \cite{voiculescu92}.

In \fig{fig:p64}, the simulated distributions match the semicircle distribution well for $D = 64$ and $512$ modes. 
For a fiber having $D = 16$ modes, the distribution is close to a semicircle distribution, but has an obvious periodic structure with 16 peaks.
The ripples become less obvious as $D$ increases from 16 to 64 to 512. 
Upon close examination of the curve for $D = 64$, the ripples seem periodic, similar to those in MMF having $D=16$ modes, but much smaller. 

The semicircle distribution, describing the GDs in a MMF with an infinite number of modes, has strict upper and lower limits, and thus a strictly bounded GD spread. 
In designing systems for MMF with a finite but large number of modes $D$, it will be sufficient to provide a GD tolerance just slightly larger than the maximum GD spread of the semicircle distribution, which is given by $4 \sigma_\mathrm{gd}$.

The GD relationship \eqn{eq:sigmamd} remains valid when the number of modes $D$ is vary large.
With a large number of modes, the relationship $\sigma_\mathrm{gd}^2 = K \sigma_\tau^2$ can be derived directly from free probability theory.

\section{Discussion}
\label{sec:dis}

The scaling of modal dispersion with fiber length in MMF is similar to the scaling of PMD in SMF. 
In a MMF shorter than the correlation length over which the local PMs can be considered constant, the GD increases linearly with fiber length. 
By contrast, in a MMF much longer than the correlation length, the number of independent sections $K$ is large, and strong mode coupling can be assumed. 
In the strong-coupling regime, a parameter describing GD per unit length may be defined as  $\sigma_\mathrm{km} = \sigma_\tau/\sqrt{L_s}$, where $L_s$ is the fiber length per section, 
\revise{2.9}{measured in kilometers.}
The overall GD, if characterized by $\sigma_\mathrm{gd}$ given by \eqn{eq:sigmamd}, is equal to $\sqrt{L_t}  \sigma_\mathrm{km}$, where $L_t = K L_s$ is the total fiber length.

In practice, there are  advantages to introducing strong mode coupling in order to reduce the modal delay spread. 
In direct-detection systems, this can reduce intersymbol interference, whereas in systems using coherent detection, this can reduce the temporal memory required in digital compensation of modal dispersion.
Recent MDM experiments \cite{li11, ryf11, salsi11}, performed in short spans of MMF, were probably not in the strong-coupling regime.
Future long-distance systems are likely to be operated in the strong-coupling regime, especially if strong mode coupling is used to reduce the overall GD spread.
\revise{1.d}{
In MMF, spatial mode coupling is governed, in part, by mode groups \cite{petar03}. 
Typically, coupling between modes in different groups is weak, with coupling length as long as 25 km \cite{kitayama80}, while coupling between modes in the same group is strong, with coupling length less than 1 km \cite{berdague82}. 
In order to reduce the GD spread in MMF, coupling between modes in different groups should be enhanced. In manufacturing of SMF, spinning is used to reduce the polarization coupling length below 100 m, thereby reducing the DGD due to PMD \cite{li98}. 
Manufacturing processes for MMF may perhaps be modified to increase spatial mode coupling in order to reduce the GD spread.
}

As seen in \figs{fig:p567} and \ref{fig:p64}(a), in the marginal p.d.f.~of GD, the number of peaks is the same as the number of modes, and the separation betweens adjacent peaks (relative to the semicircle radius) decreases with an increasing of number of modes.
In the absence of the zero-trace constraint  \eqn{zerotr_lambda}, ripples can be observed observed in Gaussian unitary ensembles up to at least $51 \times 51$ \cite[Fig. 6.1]{mehta}.
With a zero-trace constraint, the ripples are larger than those without the constraint.
In \fig{fig:p64}(b) with $D = 64$ modes, ripples are observable and seems to be very regular. 
As the number of modes increases, the ripples becomes narrower, similar to the Gibbs phenomenon \cite{hewitt79, gottlien97} for Fourier series.

Higher-order modal dispersion effects are outside the scope of this paper.
In higher-order modal dispersion, the PMs and their GDs can vary with frequency \cite{shemirani09a}. 
These effects are analogous to polarization-dependent chromatic dispersion and depolarization observed in SMF with PMD \cite{foschini00, foschini01}.
In the case of SMF with PMD, the properties of PMD to arbitrary order depend on a single parameter. 
In the case of MMF with modal dispersion and strong coupling, the higher-order properties of modal dispersion depend only on the number of modes and a single parameter, which may be taken to be the GD standard deviation $\sigma_\mathrm{gd}$ given by \eqn{eq:sigmamd}. 
%

\revise{1.c}{
In this paper, we have studied the distribution of GDs, but not the impulse response of a MMF. 
At a single frequency, the impulse response of a $D$-mode fiber consists of $D$ narrow pulses with GDs described by the distribution \eqn{jpdf_zerotr} \cite{shemirani09}, and with weights depending on the PMs excited by the transmitter launch conditions. 
Considering a modulated signal occupying a finite bandwidth, because of higher-order effects \cite{shemirani09a}, those $D$ narrow pulses broaden and may merge with each other.
The overall duration of the impulse response is described by the duration of the p.d.f.~of the GD, as shown \figs{fig:p3} to \ref{fig:p64}. 
In a fiber with many modes, where the p.d.f.~is the semicircle distribution shown in \fig{fig:p64}, the impulse response duration is just $4\sigma_{gd}$.
%
%
}

\section{Conclusion}
\label{sec:con}

In the regime of strong mode coupling, a MMF may be modeled as the cascade of \revise{2.10}{{\em many}} independent sections, which are described by statistically independent random matrices. 
The GDs are given by the eigenvalues of Gaussian unitary ensemble with zero-trace constraint.
 Marginal p.d.f.'s~of the GDs in fibers with two to seven modes have been derived analytically. 
Numerical simulations of the p.d.f.'s are in excellent agreement with analytical results. 
In a fiber with many modes, the GD is shown to follow a semicircle distribution from free probability theory. 
Numerical simulations have been conducted for fibers having $D = 16$, $64$, and $512$ modes to compare to the semicircle distribution. 


\onecolumn
\setcounter{equation}{0}
\renewcommand{\theequation}{S.\arabic{equation}}

\section*{\Large \bf Clarification and Supplement}
\begin{center}
\bf{Keang-Po Ho}
\end{center}

The main purpose of this note is to clarify the explanations in the paper \cite{ho1111}.
The group delay statistics is also extended to multimode fiber with more than $D = 7$ modes, the limitation of the direct method in \cite{ho1111}. 

For the random Gaussian unitary matrix $\mathcal{G}$ without trace constraint, the joint probability density function (p.d.f.) of the eigenvalues is given by  \cite[Sec. 3.3]{mehta} \cite{ginibre65}
\begin{equation}
p_{\mathrm{nc}}(\mathbf{x}) = \alpha_D \prod_{D \geq i > j > 0} (x_i - x_j)^2 \exp\left( - \sum_{i = 1}^D x_i^2 \right),
\label{jpdf_everytr}
\end{equation}
where $\alpha_D$ is a constant \cite[Eq. 3.3.10]{mehta} such that the p.d.f.~integrated to unity.
Here, the notation of the eigenvalue changes to $x_i$ instead of $\lambda_i$ in the paper, consistent with \cite{mehta} and other literatures. 
The zero-trace random Gaussian unitary matrix is equivalently the statistics of $\mathcal{G} - \frac{\mathcal{I}}{D} \mathrm{tr} \mathcal{G}$ with joint p.d.f.
\begin{equation}
p_{\mathrm{zt}}(\mathbf{x}) = \beta_D \delta\left( \sum_{i = 1}^D x_i \right) \prod_{D \geq i > j > 0} (x_i - x_j)^2 \exp\left( - \sum_{i = 1}^D x_i^2 \right).
\label{jpdf_zerotr}
\end{equation}
The constant $\beta_D$ is not the same as $\alpha_D$ and can be found equal to $\beta_D = \sqrt{\pi D} \alpha_D$  in later part of this note. 

For the p.d.f.~$p_{\mathrm{nc}}(\mathbf{x})$ \eqn{jpdf_everytr} without constraint, the elements of the random Gaussian unitary matrix all have the same variance. 
For the p.d.f.~$p_{\mathrm{zt}}(\mathbf{x})$ \eqn{jpdf_zerotr} with zero-trace constraint, the diagonal elements are a factor $1-1/D$ smaller in variance than other off-diagonal elements.
The combined variance of all elements of $\mathcal{G} - \frac{\mathcal{I}}{D} \mathrm{tr} \mathcal{G}$ is a factor $1 - 1/D^2$ smaller than that for $\mathcal{G}$ without constraint.
The eigenvalue variance of  $p_{\mathrm{zt}}(\mathbf{x})$  \eqn{jpdf_zerotr} with zero-trace constraint is a factor $1 - 1/D^2$ smaller than that of  $p_{\mathrm{nc}}(\mathbf{x})$ \eqn{jpdf_everytr} without constraint.
The derivation in \cite{ho1111} is consistent with this simple explanation but not as elegant here.

The fiber simulation of \cite{ho1111, ho1108} equivalently generates zero-trace Gaussian random matrix in which the variance of each element is not reduced by the factor $1 - 1/D^2$.
The random matrix in \cite{ho1111} is equivalently
\[
\frac{D}{\sqrt{D^2 - 1}} \left( \mathcal{G} - \frac{\mathcal{I}}{D} \mathrm{tr} \mathcal{G} \right).
\]
that the average variance in all elements is the same as $\mathcal{G}$ without constraint.
In \cite{ho1108}, the zero-trace constraint is directly in $\mathcal{G}$ without reducing the variance of the diagonal elements and the modification by $\mathcal{G} - \frac{\mathcal{I}}{D} \mathrm{tr} \mathcal{G}$ is not required..
In  \cite{ ho1111}, the p.d.f.~derived from \eqn{jpdf_zerotr} is required to scaled up by the factor of $D/\sqrt{D^2-1}$ to match the simulation results. 
In \cite{ho1108}, the eigenvalue p.d.f.~is pre-scaled to unity variance.

The eigenvalue distribution for zero-trace Gaussian unitary ensemble is derived according to
\begin{equation}
p_D(x_1) = \beta_D \int_{-\infty}^{+\infty} \cdots \int_{-\infty}^{+\infty} \delta\left( \sum_{i = 1}^D x_i \right) \prod_{D \geq i > j > 0} (x_i - x_j)^2 \exp\left( - \sum_{i = 1}^D x_i^2 \right)
 \ud x_2 \cdots \ud x_D.
 \label{p_D_x_1}
\end{equation}

First of all, $D \times D$ Vandemonde determinant gives
\begin{equation}
\det\left[ x_i^{j-1} \right]_{i, j = 1, 2, \dots D} =  \prod_{D \geq i > j > 0} (x_i - x_j),
\end{equation}
where $\det[ \cdot ]$ denotes a  determinant. 
Follow the method of \cite[Ch. 4]{mehta} and directly from the properties of determinant, the Vandemonde determinant can be expressed by the Hermite polynomial as
\begin{equation}
\det\left[ x_i^{j-1} \right]_{i, j = 1, 2, \dots D} =  \det\left[ 2^{-(j-1)} H_{j-1}(x_i) \right]_{i, j = 1, 2, \dots D} 
\end{equation}
and
\begin{equation}
\det\left[ x_i^{j-1} \right]_{i, j = 1, 2, \dots D}= \det\left[ 2^{-(j-1)} H_{j-1}(x_i + c_i ) \right]_{i, j = 1, 2, \dots D},
\label{vendershift}
\end{equation}
where $H_n(x)$ is the Hermite polynomial as given in \cite{mehta} and $c_i$ is constant. 
The leading terms in both $ 2^{-(j-1)} H_{j-1}(x_i)$ and $2^{-(j-1)} H_{j-1}(x_i + c_i )$ are $x_i^{j-1}$.
Using the Hermite polynomial, the p.d.f. \eqn{jpdf_everytr} can be expressed as \cite{mehta}

\begin{equation}
 p_{\mathrm{nc}}(\mathbf{x}) =  \frac{1}{D} \det\left[ K_D(x_i, x_j) \right]_{i, j = 1, 2, \dots D},
  \label{Det_notr}
\end{equation}
where
\begin{equation}
K_D(x, y) = \sum_{k = 0}^{D-1} \frac{1}{2^k k! \sqrt{\pi}} H_k(x) H_k(y) e^{-x^2/2 - y^2/2}.
\label{KDxy}
\end{equation}
Without constraint, the eigenvalue distribution is given by $\frac{1}{D} K_D(x, x)$
and $K_D(x, x)$ is called correlation function \cite{mehta}.

With zero-trace constraint, the p.d.f. \eqn{jpdf_zerotr} becomes
\begin{equation}
p_{\mathrm{zt}}(\mathbf{x}) = \frac{\beta_D}{D \alpha_D} \delta\left( \sum_{i = 1}^D x_i \right) \det\left[ K_D(x_i, x_j) \right]_{i, j = 1, 2, \dots D}.
\end{equation}

Similar to \cite{majumdar08, chen10} but using Fourier instead of Laplace transform, 
\begin{equation}
p_{\mathrm{zt}}(\mathbf{x}) = \frac{\beta_D}{2 \pi D \alpha_D} \int_{-\infty}^{+\infty} \det\left[ K_D(x_i, x_j) \right]_{i, j = 1, 2, \dots D} \exp\left(i \omega \sum_{i = 1}^D x_i \right) \ud \omega,
\end{equation}
or
\begin{equation}
p_{\mathrm{zt}}(\mathbf{x}) = \frac{\beta_D}{2 \pi D \alpha_D} \int_{-\infty}^{+\infty} \det\left[ K_D\left( x_i + \frac{i\omega}{2}  , x_j + \frac{i\omega}{2}   \right) \right]_{i, j = 1, 2, \dots D} \exp \left( - \frac{D}{4} \omega^2 \right) \ud \omega
\label{omega_in}
\end{equation}
In \eqn{omega_in}, the argument for the Hermite polynomial in \eqn{KDxy} changes from $x_i$ to $x_i + {i\omega}/{2} $ using \eqn{vendershift}. 

Similar to the integration of \eqn{Det_notr} to obtain the eigenvalue distribution, the p.d.f.~\eqn{p_D_x_1} is given by
\begin{equation}
p_D(x) = \frac{\beta_D}{2 \pi D \alpha_D} \int_{-\infty}^{+\infty} K_D\left( x + \frac{i\omega}{2} , x + \frac{i\omega}{2}  \right)  \exp \left( - \frac{D}{4} \omega^2 \right) \ud \omega,
\end{equation}
or
\begin{equation}
p_D(x) = \frac{\beta_D}{2 \pi D \alpha_D}      \int_{-\infty}^{+\infty}
                \exp \left[ - \frac{D}{4} \omega^2 - \left( x + \frac{i\omega}{2} \right)^2 \right]            
                \sum_{k = 0}^{D-1} \frac{H_k^2 \left( x +\frac{1}{2} i\omega \right)}{2^k k! \sqrt{\pi}}   
                \ud \omega.
   \label{p_D_x_int1}
\end{equation}
In the integration \eqn{p_D_x_int1}, we may conduct the first integration over $x$ before integration over $\omega$ to obtain $\beta_D/\alpha_D = \sqrt{\pi D}$. 
The integration  $p_D(x)$ \eqn{p_D_x_int1} becomes
\begin{equation}
p_D(x) = \frac{1}{2 \sqrt{\pi D} }      \int_{-\infty}^{+\infty}
                \exp \left[ - \frac{D}{4} \omega^2 - \left( x + \frac{i}{2} \omega \right)^2 \right]            
                \sum_{k = 0}^{D-1} \frac{H_k^2 \left( x + \frac{i}{2} \omega \right)}{2^k k! \sqrt{\pi}}   
                \ud \omega.
   \label{p_D_x_int}
\end{equation}

Using the substitute of $s = i \omega$, the integration \eqn{p_D_x_int} is very similar to the Mellin's inverse formula for Laplace transform with an integration from  $x - i \infty$ to $x + i \infty$.
The following algebraic expression can be obtained based on Laplace transform

\begin{equation}
p_D(x) = \frac{ \exp\left( -\frac{D}{D-1} x^2 \right) }
           {\sqrt{\pi D (D-1)}}
           \sum_{k = 0}^{D-1} \frac{1}{2^k k!} \left.  H_k^2 \left( \frac{t}{2 \sqrt{D-1}} \right)
          	\right| _{t^n \leftarrow (-1)^n H_n \left(  \frac{D x}{\sqrt{D-1}}  \right)}
          	\label{p_D_sub}
\end{equation}
In \eqn{p_D_sub}, the expression inside the summation is a $2(D-1)$ degree polynomial of $t$.
The power $t^n$ is replaced by the Hermite polynomial of $(-1)^n H_n \left(  \frac{D x}{\sqrt{D-1}} \right)$.
The derivation of \eqn{p_D_sub} is based on the Laplace transform relation of $\mathcal{L}\left[ \ud^n f/\ud t^n \right] = s^n \mathcal{L}[f]$ and $\ud^n e^{-x^2}/\ud x^n = (-1)^n H_n(x) e^{-x^2}$, where $\mathcal{L}$ denotes the Laplace transform. 

\begin{figure}[t]
\centering{
 \includegraphics[width = 0.45 \textwidth]{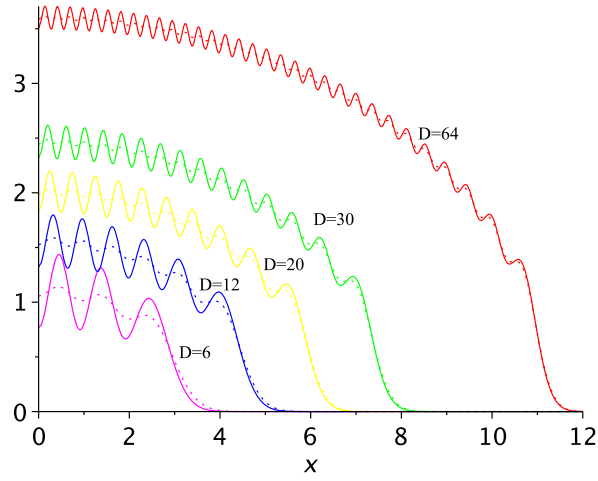}}
 \caption{
 The distribution of $D p_D(x)$ (solid curves) for zero-trace Gaussian unitary ensemble as compared with 
 the correlation function of $K_D(x,x)$ (dot curves) without constraint. 
 }
\label{fig:pdf}
\end{figure}
 \Fig{fig:pdf} shows the distribution of $D \times p_D(x)$, corresponding to the correlation function in \cite{mehta}, as compared with $K_D(x, x)$ without constraint.
 The ripple in $D p_D(x)$ is far larger than the corresponding function without zero-trace constraint.
 With the increase of dimension $D$, the distribution approaches semicircle distribution, similar to the conclusion in \cite{ho1111}.
   
In \cite{tracy01}, the monotone sequences in random words are found to have the same statistics as zero-trace Gaussian unitary ensemble.
Using the notation here, $p_D(x)$  is given by the integration equation\footnote{This argument is from Prof. Folkmar Bornemann of Tech. Univ. Munich, see \url{http://mathoverflow.net/questions/86965/traceless-gue-four-centered-fermions}}:
\begin{equation}
\frac{1}{D} K_D(x, x) = \sqrt{ \frac{D}{\pi}} \int_{-\infty}^{+\infty} e^{-D y^2} p_D(x - y) \ud y
\label{p_D_int_eq}
\end{equation}
The original integration equation \eqn{p_D_int_eq} is for the largest eigenvalue but the same argument applies to $p_D(x)$. 
The left hand side of \eqn{p_D_int_eq} is the eigenvalue distribution of Gaussian unitary ensemble without constraint. 
Using Fourier transform to both side of \eqn{p_D_int_eq}, $p_D(x)$ can be found and should be the same as \eqn{p_D_sub}.

 In summary, we have extended the p.d.f.~for zero-trace Gaussian unitary ensemble to very high order.

\end{document}